\documentclass[usenatbib,useAMS,usegraphicx]{mn2e}

 \usepackage{times}
 \usepackage{amsmath}
 \usepackage[latin1]{inputenc}
 \usepackage[usenames,dvipsnames]{color}

  \definecolor{brown}{RGB}{165,42,42}

  \newcommand{\kms}   {~km~s$^{-1}$}
  \newcommand{\mjy}   {~mJy~beam$^{-1}$}
  \newcommand{\jy}    {~Jy~beam$^{-1}$}
  \newcommand{\cmd}   {~cm$^{-2}$}
  \newcommand{\cmt}   {~cm$^{-3}$}
  \newcommand{\vlsr}  {$v_{\rm LSR}$}
  \newcommand{\mo}    {$M_{\sun}$}
  \newcommand{\tdust} {T_\mathrm{dust}}
  \newcommand{\Tex}   {T_\mathrm{ex}}
  \newcommand{\jr}[2] {\mbox{$J$=#1$\rightarrow$#2}}
  \newcommand{\jt}[2] {\mbox{#1$\rightarrow$#2}}
  \newcommand{\arcdeg}{\mbox{$^\circ$}} 
  \newcommand{\phn}   {\phantom{0}}     
  \newcommand{\hcop} {HCO$^+$}
  \newcommand{\hdco} {H$_2$CO}
  \newcommand{\ndh}  {N$_2$H$^+$}
  \newcommand{\nht}  {NH$_3$}
  \newcommand{\cts}  {C$^{34}$S}
  \newcommand{\htcop}{H$^{13}$CO$^+$}
  \newcommand{\cthd} {$c$-C$_3$H$_2$}
  \newcommand{\jk}[2]{$#1_{#2}$} 
  \newcommand{\jj}[4]{\jk{#1}{#2}--\jk{#3}{#4}}
  \newcommand{\jf}[4]{$\frac{#1}{#2}$,$\frac{#3}{#4}$} 
  \newcommand{\fj}[4]{$\frac{#1}{#2}$--$\frac{#3}{#4}$}

\begin{document}

\title[Probing the physical and chemical structure of the CS core in LDN~673]
  {Probing the physical and chemical structure of the CS core in
  LDN~673. Multitransitional and continuum observations}

\author[Morata et al.]{
 Oscar Morata$^{1}$, Josep Miquel Girart$^{2}$,  Robert Estalella$^{3}$ and
 Robin T. Garrod$^{4}$ \\
  $^{1}$Institute of Astronomy \& Astrophysics, Academia Sinica, P.O.\ Box
   23-141, Taipei 10617, Taiwan; omorata@asiaa.sinica.edu.tw \\ 
  $^{2}$Institut de Ci\`encies de l'Espai (CSIC-IEEC), Campus UAB, Facultat de
   Ci\`encies, C5 par 2$^a$, E-08193, Bellaterra, Catalunya, Spain \\
  $^{3}$Departament d'Astronomia i Meteorologia (IEEC-UB), Institut de
   Ci\`encies del Cosmos, Universitat de Barcelona, Mart\'{i} i Franqu\`es 1,
   E-08028 Barcelona,\\ Catalunya, Spain \\
  $^{4}$Department of Astronomy, Cornell University, Ithaca, NY 14853, USA\\}

\maketitle

\begin{abstract}
  High-angular resolution observations of dense molecular cores show that
  these cores can be clumpier at smaller scales, and that some of these clumps
  can also be unbound or transient. The use of chemical models of the
  evolution of the molecular gas provides a way to probe the physical
  properties of the clouds.  We study the properties of the clump and
  inter-clump medium in the starless CS core in LDN 673 by carrying out a
  molecular line survey with the IRAM 30-m telescope toward two clumps and two
  inter-clump positions. We also observed the 1.2-mm continuum with the
  MAMBO-II bolometer at IRAM.  The dust continuum map shows four
  condensations, three of them centrally peaked, coinciding with previously
  identified sub-millimetre sources.  We confirm that the denser clump of the
  region, $n\sim3.6 \times10^5$\cmt, is also the more chemically evolved, and
  it could still undergo further fragmentation. The inter-clump medium
  positions are denser than previously expected, likely
  $n\sim1\times10^3$--1$\times10^4$\cmt\ due to contamination, and are
  chemically young, similar to the gas in the lower density clump position.
  We argue that the density contrast between these positions and their general
  young chemical age would support the existence of transient clumps in the
  lower density material of the core. We were also able to find reasonable
  fits of the observationally derived chemical abundances to models of the
  chemistry of transient clumps.
\end{abstract}

 \begin{keywords}
  ISM: individual objects: LDN 673 --- ISM: abundances --- ISM:
  clouds --- ISM: molecules --- radio lines: ISM --- stars: formation
 \end{keywords}

\section{Introduction}
 \label{intro}

  It has been long known that molecular clouds are highly structured
  \citep[e.\, g.,][]{BlitzStark86} and their structure is greatly affected by
  the motions induced by supersonic turbulence \citep[e.\ g.,][]{Scalo98},
  self-gravity of the gas and magnetic fields inside the clouds
  \citep{McKeeOstriker07,Kainulainen09}. All these processes control the
  formation and evolution of the density enhancements, of different scale
  sizes and densities, such as the cores and clumps that will finally give
  birth to stars.  But there are still great uncertainties to identify the
  connection between starless cores and protostars
  \citep{Johnstone00,SmithClark08}. Starless cores are not all the same,
  despite their overall similarity in structure \citep{KetoCaselli08}, and
  differences in total mass, density and temperature might account for the
  differences in dynamical properties, structure, and future evolution of
  starless cores \citep{KetoField05,KetoCaselli08}.

  Higher-angular resolution observations are also finding that dense cores in
  molecular clouds that appeared homogeneous in single-dish observations, are
  clumpier at smaller scales \citep{Peng98,Morata03}, showing structures as
  small as 0.02 pc, and masses as low as 0.01\mo. Many of these smaller clumps
  are unbound and/or showing evidence of being transient
  \citep{Peng98,Morata05}, and will never be able to form low-mass stars or
  even brown dwarfs. The mix of bound and unbound structures in dense cores is
  also found in several regions, such as the Pipe Nebula, where recent
  molecular line and continuum observations
  \citep{Lombardi06,Muench07,Rathborne08,Frau10} found numerous cores more
  than 100, most of which appear to be pressure confined, and gravitationally
  unbound \citep{Lada08}.

  Observations of the emission of molecular lines at millimetre and
  sub-millimetre wavelengths in cloud cores combined with the modelling of the
  chemistry of the gas provides a way of obtaining information on the physical
  structure and the chemical and physical evolutionary stages of the cores. We
  proposed a time-dependent chemical model that also explored the consequences
  of the presence of unresolved and transient structures in the gas that would
  form and disperse in a timescale of $\sim1$--2 Myrs \citep{Taylor96}, in
  order to explain the systematic differences between CS and \nht\ lines
  \citep{Pastor91,Morata97}.  Simulations of the evolution of these cores
  \citep{Garrod05,Garrod06} find that clouds that are ensembles of such
  transients have a clearly different chemistry from a `traditional' static
  cloud. The gas chemistry appears to be ``young'' at all times, and the
  re-cycling of the material frozen out onto dust grains produces a general
  molecular enrichment of the clouds, even after re-expansion of the transient
  structures. The background gas in which these inhomogeneities are embedded
  would be fairly diffuse, but chemically enriched. These chemical
  enhancements might also account for the variety of chemistries observed in
  diffuse clouds.

  We carried out interferometric high-angular resolution observations of
  several molecules (CS, \hcop\, and \ndh) \citep[][hereafter
    MGE03]{Morata03}, which we later combined with single-dish
  intermediate-angular resolution maps \citep[][from now on MGE05]{Morata05}
  towards the starless CS core in LDN~673 ($d=300$~pc), in order to test the
  predictions of the chemical models. The combined single-dish and
  interferometer maps showed emission of both background and clumped gas, with
  a clear segregation of clump properties between the northern and southern
  halves of our observed region, and allowed us to identify 15 resolved clumps
  in our data cube. The derived clump masses are well below the virial mass,
  which would point to their being transient, except for the more massive one,
  which might have a mass, $\sim 1$\mo, closer to the virial mass.  The
  starless core appears to be constituted by a heterogeneous medium of
  condensations, of various densities and at different stages of chemical
  evolution, in agreement with theoretical studies that postulate the
  existence of transient clumps or the transient nature of dense cores
  generated by dynamical flows within molecular clouds \citep[see e.\,
    g.,][]{FH02, Vazquez05, VanLoo08}. Recently, \citet{Whyatt10} also found
  evidence of a heterogeneous medium in scales of less than 0.1 pc near HH
  objects, as traced by strong \hcop\ (3-2) emission. These clumps would have
  gas volume densities $\ga3\times10^4$\cmt.

 \begin{table}
 \caption{Lines observed with the 30-m IRAM telescope}
 \label{tablines}
   \centering

   \begin{tabular}{lccc}
     \hline\noalign{\smallskip}
     Transition                     & Frequency & HPBW & $B_\mathrm{eff}$ \\ 
                                    & (GHz)     & (arcsec)  &\\
     \noalign{\smallskip}\hline\noalign{\smallskip}
     CCH 1--0 \fj{3}{2}{1}{2} $F$=2--1        & \phn87.316925 &    28 & 0.78 \\
     CCH 2--1 \fj{5}{2}{3}{2} $F$=3--2        & 174.663222    &    14 & 0.64 \\
     CCH 3--2 \fj{7}{2}{5}{2} $F$=4--3        & 262.004260    & \phn9 & 0.46 \\
     \noalign{\smallskip}
     CN 1--0 \jf{3}{2}{5}{2}--\jf{1}{2}{3}{2} & 113.490982    &    22 & 0.74 \\
     CN 2--1 \jf{5}{2}{7}{2}--\jf{3}{2}{5}{2} & 226.874764    &    11 & 0.54 \\
     \noalign{\smallskip}
     CS 5--4                                  & 244.935606    &    10 & 0.50 \\
     \noalign{\smallskip}
     \cthd\ \jj{2}{1,2}{1}{0,1}               & \phn85.338906 &    29 & 0.78 \\
     \cthd\ \jj{4}{1,4}{3}{0,3}		   & 150.851899    &    16 & 0.68 \\
     \noalign{\smallskip}
     HCN 3--2				   & 265.886432    & \phn9 & 0.45 \\
     \noalign{\smallskip}
     \hdco\ \jj{2}{1,2}{1}{1,1}		   & 140.839515    &    18 & 0.70 \\
     \hdco\ \jj{3}{1,3}{2}{1,2}            & 211.211448    & 12 & 0.57 \\
     \hdco\ \jj{3}{1,2}{2}{1,1} 	   & 225.697773    & 11 & 0.54 \\
     \noalign{\smallskip}
     \htcop\ 1--0 			   & \phn86.754330 &    28 & 0.78 \\
     \htcop\ 2--1 			   &    173.506782 &    14 & 0.64 \\
     \htcop\ 3--2			   &    260.255480 &    10 & 0.46 \\
     \noalign{\smallskip}
     NO \jf{3}{2}{5}{2}--\jf{1}{2}{3}{2} $\Pi^+$ & 150.176459 &    16 & 0.68 \\
     NO \jf{5}{2}{7}{2}--\jf{3}{2}{5}{2} $\Pi^+$ & 250.436845 &    10 & 0.48 \\
     \noalign{\smallskip}
     SO       \jj{3}{2}{2}{1}		   & \phn99.299905 &    25 & 0.76 \\
     SO       \jj{4}{5}{3}{4} 		   &    206.176062 &    12 & 0.58 \\
     SO       \jj{6}{5}{5}{4} 		   &    219.949433 &    11 & 0.55 \\
     SO       \jj{7}{6}{6}{5} 		   &    261.843756 &    10 & 0.46 \\
     \noalign{\smallskip}
     SO$_2$   \jj{3}{1,3}{2}{0,2} 	   &    104.029410 &    24 & 0.76 \\
     SO$_2$   \jj{7}{1,7}{6}{0,6}          &    165.225436 &    15 & 0.66 \\
     \noalign{\smallskip}\hline
   \end{tabular}
 \end{table}

 \begin{table}
 \caption{Coordinates of the positions selected for the spectral line
     observation.}
 \label{tabpositions}
   \begin{tabular}{llll}
     \hline\noalign{\smallskip}
     Position & Counterpart & R.A.(J2000) & Dec. (J2000)\\
     \noalign{\smallskip}\hline\noalign{\smallskip}
     CL1  & \ndh\ peak    & 19:20:51.747 & 11:13:49.50 \\
     CL6  & CS peak       & 19:20:50.003 & 11:14:53.00 \\
     ICLN & Inter-clump N & 19:20:51.701 & 11:15:30.00 \\
     ICLS & Inter-clump S & 19:20:54.501 & 11:14:10.00 \\
     \noalign{\smallskip}\hline
   \end{tabular} 
 \end{table}

  In order to study the properties of the clump and inter-clump gas in the
  starless CS core in LDN 673, we selected two positions associated with
  identified clumps (CL1 and CL6) and two positions where the inter-clump gas
  would be dominant (where we did not detect any clump). A multitransitional
  survey of several early- and late- type molecules in these positions allows
  us to sample the chemical composition of the gas and compare it to the
  predicted different chemistry of the pre- and post-clump gas in the
  models. We additionally observed the dust continuum emission in LDN~673. The
  structure of this paper is as follows: in Sect.~\ref{observations}, we
  describe the IRAM 30-m spectral line and continuum observations. In
  Sect.~\ref{results}, we describe the characteristics of the detected spectra
  and of the dust continuum emission. The analysis of the observational
  results and the determination of the physical parameters of the gas and dust
  are shown in Sect.~\ref{analysis}.  Finally, Sect.~\ref{discussion} contains
  the discussion of the results of our analysis and how they can be related to
  the previous observations, the chemistry of the clouds and the structure of
  the core.

\section{Observations}
 \label{observations}

  The spectral line observations were carried out in 2005 August using the
  30-m IRAM telescope in Granada (Spain). We used the capability of the ABCD
  multi-receiver system to observe 10 different molecules (and a total of 23
  transitions) with just 6 frequency setups in the 3, 2, and 1-mm
  bands. Table~\ref{tablines} shows the transitions and frequencies observed.
  We used the VESPA autocorrelator as a spectral backend, which provided a
  total bandwidth of 80 MHz, and selected a 20 kHz channel spacing for the
  receiver at 100 GHz, and a 40 kHz channel spacing for the other three
  receivers. The achieved velocity resolutions range from 0.4 to
  0.07\kms\ from the 1 to the 3-mm bands. The main-beam efficiencies and the
  half-power beam widths at the observed frequencies are also listed in
  Table~\ref{tablines}. We used the frequency-switching mode with a 7.9 MHz
  throw, except for the configuration that observed simultaneously the CCH
  (1--0), \cthd\ (\jj{4}{1,4}{3}{0,3}), \hdco\ (\jj{3}{1,2}{2}{1,1}), and SO
  (\jj{7}{6}{6}{5}) transitions, where we used a 15.8 MHz throw. We obtained
  system temperatures, in $T_\textrm{a}^*$ scale, of 85-160~K at 100~GHz,
  215--465~K at 150~GHz, 230--320~K at 230~GHz, and 470--680~K at 270~GHz. We
  used the GILDAS\footnote{\tt http://www.iram.fr/IRAMFR/GILDAS} package
  of IRAM to reduce, analyse, and display the spectral data.

 \begin{figure}
   \begin{center}
     \includegraphics[angle=-90,width=\columnwidth]{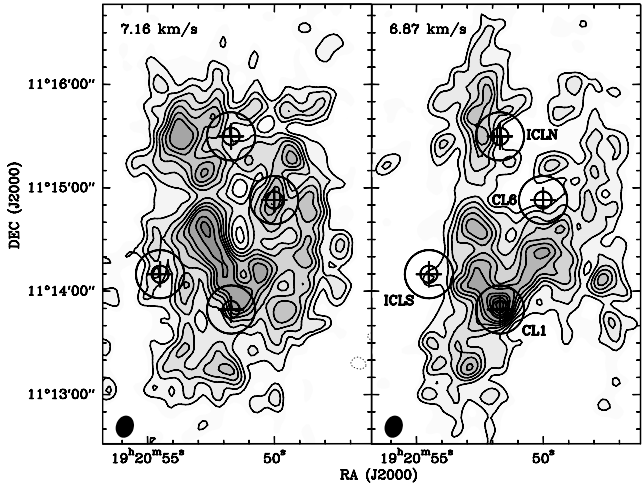}
   \end{center}
 \caption{Positions observed with the 30-m telescope in the spectral line
   observation \textit{(crosses)} overlapped with the greyscale and contour
   channel map of the CS (2--1) emission obtained by MGE05. The two concentric
   circles mark the approximate 30-m beam sizes at 1 and 3 mm, respectively.}
 \label{fobspos}
 \end{figure}

  Table~\ref{tabpositions} gives the coordinates and names of the four
  observed positions. Figure~\ref{fobspos} plots the location of the observed
  positions over the combined BIMA-FCRAO map of the CS (\jr{2}{1}) emission of
  MGE05. The four positions were selected according to the following criteria:
  (1) two positions centred in two of the CS clumps detected by MGE05: CL1,
  the most massive and chemically evolved clump, and CL6, a chemically young
  clump found in the northern region of the BIMA map; (2) two inter-clump
  positions, one in the northern region (ICLN) and another in the southern
  region (ICLS), selected from the study of clumpiness of the CS emission in
  MGE05.

 \begin{figure}
   \begin{center}
     \includegraphics[width=\columnwidth]{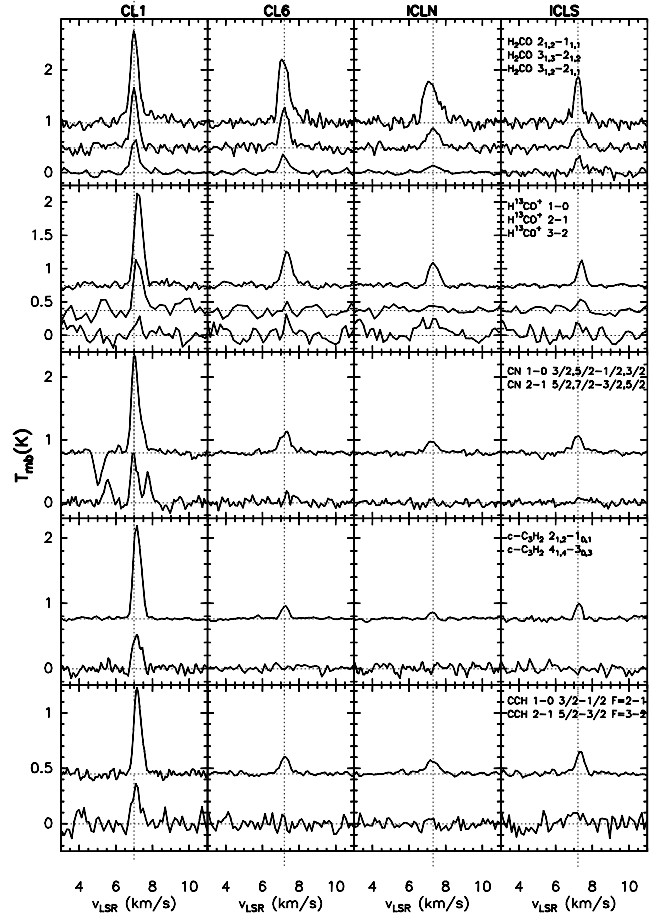}
   \end{center}
 \caption{Spectra of the detected transitions of the \hdco, \htcop, CN, \cthd,
   and CCH molecules at the four selected positions in the CS core of LDN 673,
   in the velocity range 3--11\kms. The dotted vertical lines indicate the
   $V_\mathrm{LSR}$ of the \hdco\ (\jj{3}{1,3}{2}{1,2}) transition for the
   respective position.}
 \label{figspectra1}
 \end{figure}

 \begin{figure}
   \begin{center}
     \includegraphics[width=\columnwidth]{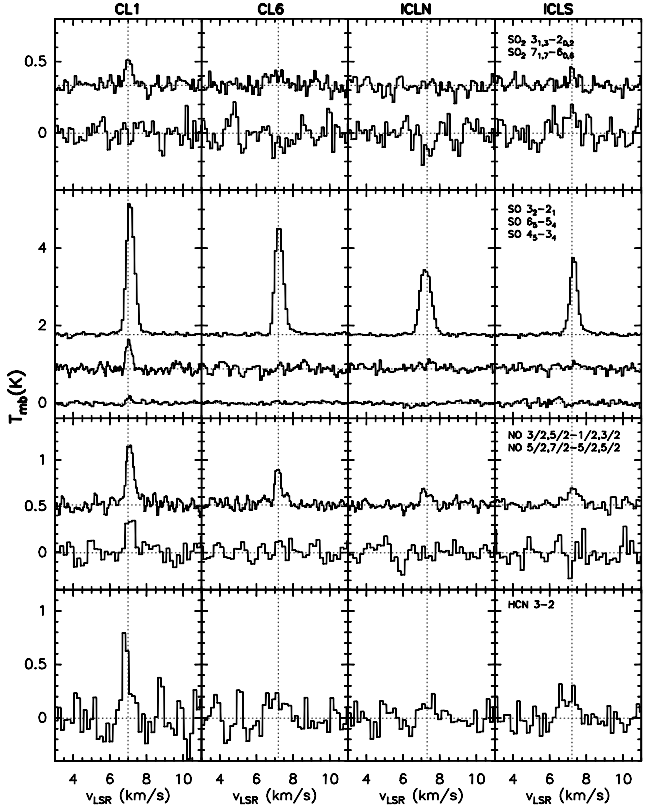}
   \end{center}
 \caption{Same as Fig. \ref{figspectra1} for the SO, SO$_2$, NO, and HCN
   molecules.}
 \label{figspectra2}
 \end{figure}

 \begin{figure}
   \begin{center}
     \includegraphics[width=\columnwidth]{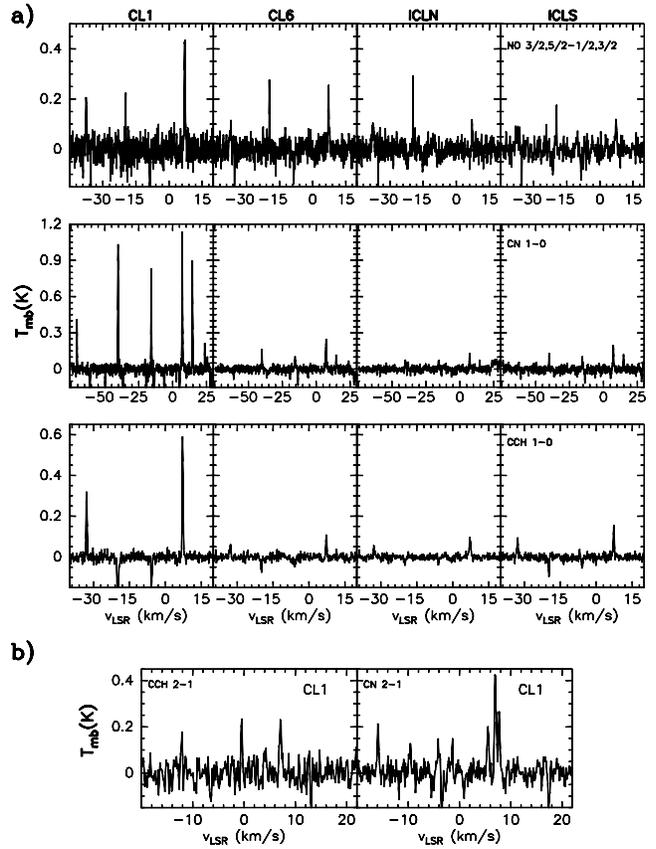}
   \end{center}
 \caption{Spectra of the detected lines showing hyperfine structure:
   \textit{a)} CN, CCH and NO lines at 3-mm; \textit{b)} CCH (2--1) and CN
   (2--1) lines in CL1.}
 \label{fighfs}
 \end{figure}

  The continuum observations were carried out in Feb\-ruary 2007 with the 30-m
  telescope using the MAMBO-II 117-channel bolometer at 1.2~mm. The bolometer
  effective frequency is 250 GHz and the half-power beam-width at 1.2-mm is
  $11''$. The source was observed using the On-the-fly mode, with a secondary
  chopping of $42''$. The telescope was scanning in azimuth at a speed of
  $8''$~s$^{-1}$. Total observation time was 1.4~h, with 57 minutes on source.
  During the observation, the zenith atmospheric opacity at 225~GHZ was 0.32.
  The resulting rms was $\simeq6$\mjy.  The map was centred at $\alpha
  \mathrm{(J2000)} = 19^\mathrm{h}20^\mathrm{m}52\fs20$; $\delta
  \mathrm{(J2000)} = 11\arcdeg14'45\farcs5$ and it covers an effective area of
  about $9\farcm0\times6\farcm5$ along the NE--SW direction. We used the
  MOPSIC package of IRAM to reduce the continuum data and the MIRIAD
  \citep{Sault95} and GILDAS software package to analyse and display the
  continuum data.

\section{Results}
 \label{results}

\subsection{Spectral line observations}

  Figures~\ref{figspectra1} and \ref{figspectra2} show the spectra obtained in
  the four selected positions for the detected molecules (19 of the 23 lines
  observed). Tables~\ref{tline_cl1} to \ref{tline_icls} show the line
  parameters obtained using a Gaussian fit for the transitions detected in the
  CL1, CL6, ICLN, and ICLS positions, respectively.  Table~\ref{tline_hfs}
  shows the hyperfine structure fit parameters (obtained using the HFS method
  of the CLASS package) for the detected CN, CCH and NO transitions
  (Figure~\ref{fighfs} shows the full spectra of all the detected hyperfine
  structure lines).  Finally, Table~\ref{tline_nd} gives the upper limits for
  the line intensities of the transitions not detected at each position. The
  CCH (\jt{3}{2}), CS (\jt{5}{4}), SO (\jt{\jk{7}{6}}{\jk{6}{5}}), and SO$_2$
  \jj{7}{1,7}{6}{0,6} lines were not detected at any position. The first three
  lie in the 1-mm band, which presents higher $T_\textrm{sys}$, and typically
  shows a higher rms than the 3- and 2-mm band values in our spectra.

  All the 19 detected transitions are at the CL1 position, whereas the other
  three positions show detection in 10 (CL6 and ICLN) or 11 (ICLS)
  transitions.  If we compare the intensity for the same line across the four
  positions, the emission is always more intense at the CL1 position. Lines at
  the CL6 position tend to be more intense than at ICLS, while lines at the
  ICLN position are typically the less intense ones.

  There are not important differences in the line central velocities among the
  four positions. The four observed positions show \vlsr\ velocity differences
  of 0.1--0.2\kms\ between them, and are in agreement with the velocity
  pattern found by MGE05.  Linewidths toward CL1 tend to be the narrowest
  ($\sim0.4$--0.5\kms, typically). This suggests that the turbulence is lower
  in CL1 than in the rest of the region.

\subsection{Dust continuum emission}

\begin{figure}
   \begin{center}
     \includegraphics[angle=-90,width=\columnwidth]{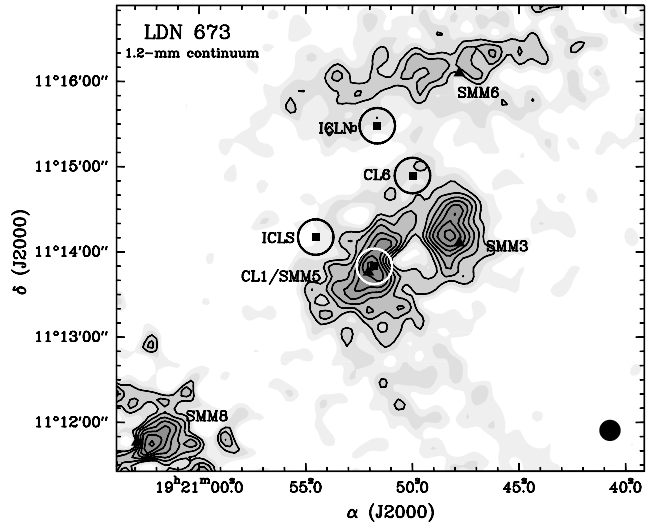}
   \end{center}
 \caption{1.2-mm dust continuum map obtained with the MAMBO-II bolometer after
   convolving with a $15''$ Gaussian beam. The rms noise of the map is
   7\mjy. The minimum contour is 21\mjy, with increments of 7\mjy.  The
   circles show the beam size of the 30-m telescope at 3mm, $\sim25''$,
   centred at each of the positions of the spectral line survey (solid
   squares). The solid triangles mark the positions of the SMM sources found
   by \citet{Visser02}.}
 \label{fmambo}
 \end{figure}

  Figure~\ref{fmambo} shows the 1.2-mm continuum map obtained with the
  MAMBO-II bolometer. We find four emission condensations with different
  degrees of extension and intensity.  These four condensations were
  previously detected at 850~$\mu$m by \citet{Visser02}. Table~\ref{tdustcond}
  gives the emission parameters derived for these four condensations and the
  association with the sources of \citet{Visser02}. These four condensations
  were also recently identified by \citet{Tsitali10} using the \textit{Spitzer
  Space Telescope}, which they classify as starless cores.

  The three southern condensations show an approximately rounded shape and
  with centrally peaked emission, whereas the northern condensation rather
  looks as an extended patch of weak emission.  The most intense emission peak
  (the central clump in Fig.~\ref{fmambo}), SMM5, coincides with CL1 and with
  the \ndh\ emission of MGE03.  Another, slightly smaller, bright condensation
  is SMM3, with a similar value of the emission peak and located about $\simeq
  50''$ west of SMM5.  A third condensation, again slightly smaller, but with
  a similar emission peak to the previous two, is SMM8, located $\sim2\farcm5$
  SE of SMM5, near the edge of the 1.2 mm observation's field of view. This
  condensation was also detected in \htcop\ (1--0) and \cts\ (2--1) in
  single-dish observations (MGE05). The fourth condensation, associated with
  SMM6, presents diffuse and extended dust emission in an approximately E--W
  strip $\sim2'$ north of SMM5 that coincides with a ridge of molecular
  emission found in an E--W direction in the maps obtained with the FCRAO
  telescope (MGE05).

  Among the four positions observed in spectral lines, only CL1 is associated
  with a dust peak matching very well with SMM5. CL6 and ICLN are associated
  with weak dust emission at $2\sigma$, and ICLS is associated with even
  weaker emission.

\section{Analysis}
 \label{analysis}

 \begin{table*}
 \label{tdustcond}
   \caption{Parameters of the condensations found in the 1.2-mm dust emission}
   \centering
   \begin{tabular}{lccccccccl}
     \hline\noalign{\smallskip}
     &	      &	           & Peak   & Flux$^a$ & && & & \\
     & \multicolumn{2}{c}{Position} & intensity & density & FWHM$^b$ &
     Mass$^c$ &  $N(\mathrm{H_2})\,$$^d$ & $n(\mathrm{H_2})\,$$^d$    & \\
     \cline{2-3}\noalign{\smallskip}
     No. & R.A.(J2000) & DEC.(J200) & (\jy) & (Jy) & (arcsec) & 
     (\mo)& ($10^{22}$\cmd) & ($10^5$\cmt) &  Association$\,^e$\\
     \noalign{\smallskip}\hline\noalign{\smallskip}
     1 & 19:20:46.0 & 11:16:20.0 & 0.038 & 0.436 & 75 & 3.9 & 1.6 & 0.7 & SMM6
     \\ 
     2 & 19:20:48.2 & 11:14:10.5 & 0.066 & 0.253 & 41 & 2.9 & 4.0 & 3.3 & SMM3
     \\ 
     3 & 19:20:52.0 & 11:13:49.5 & 0.066 & 0.339 & 49 & 5.1 & 4.9 & 3.4 &
     CL1/SMM5 \\
     4 & 19:21:02.4 & 11:11:43.5 & 0.065 & 0.186 & 35 & 1.6 & 3.0 & 2.8 & SMM8
     \\ 
     \noalign{\smallskip}\hline\noalign{\medskip}
   \end{tabular}

   \begin{minipage}{16.2cm}
     \begin{list}{}{}
       \item[$^a$] Flux density integrated above the 2$\sigma$ level after
         convolving with a 15'' Gaussian beam.
       \item[$^b$] FWHM of the dust emission condensation,
         $\theta_\mathrm{FWHM} = \sqrt{\theta_s^2 - \theta_b^2}$, where
         $\theta_b$ is the beam-size and $\theta_s$ = $2\sqrt{A /\pi}$. $A$ is
         the area on the map of the pixels with intensity over half of the
         local continuum emission peak.
       \item[$^c$] Total mass calculated from the flux density integrated
         above the 2$\sigma$ level, using the expression from \citet{Frau10},
         for a distance of 300~pc.
       \item[$^d$] $N(\mathrm{H_2})$ and $n(\mathrm{H_2})$ are the FWHM
         averaged values of column and volume densities, respectively,
         calculated using the expressions from \citet{Frau10}, assuming dust
         temperature, $\tdust=10$~K, a standard dust-to-gas ratio of 100, a
         dust absorption coefficient at a frequency of 250 GHz, $\kappa_{250}
         = 0.0065$ cm$^2$ g$^{-1}$ (taken as a medium value between dust
         grains with thin and thick ice mantles for a volume density of
         $\sim10^5$\cmt, \citet{Ossenkopf94}, which should be typical for
         starless cores.
       \item[$^e$] SMM: sub-millimetre sources from \citet{Visser02}.
     \end{list}
   \end{minipage}
 \end{table*}

\subsection{Radiative transfer analysis}
 \label{sect_radtran}

  The detection of multiple transitions at the same position of some molecules
  allows us to constrain the physical properties of the emitting gas. We used
  the RADEX radiative transfer model code \citep{Radex}, which treats optical
  depth effects with an escape probability method, to find the set of
  molecular column densities, $N(X)$, volume densities, $n$(H$_2$), and
  kinetical temperatures, $T_\mathrm{K}$, that best reproduced the observed
  lines. Appendix~\ref{radexmethod} gives more details on how the calculations
  were done.

 \begin{table*}
 \caption{Physical parameters derived from the best-fit model defined by the
   minimum value of the $\chi^2_\nu$ function of a series of RADEX models
   compared to the lines of \hdco, SO, and \htcop\ observed in the four
   positions.}
 \label{radextable}
   \centering
   \begin{tabular}{llcccccrr}
     \hline\noalign{\smallskip}
     Position & Species & $\theta_s$ & $T_K$ & $n(\mathrm{H}_2)$ &
     $N(\mathrm{X})^a$ & Transition & $\Tex^b$ & \multicolumn{1}{c}{$\tau^b$}\\
     & & ($''$) & (K) & ($10^{5}$ cm$^{-3}$) & (cm$^{-2}$) && (K) &\\
     \noalign{\smallskip}\hline\noalign{\smallskip}
     CL1  & \hdco\ & 50 & 10 & 3.6 & $8.4\times 10^{12}$ & \jj{2}{1,2}{1}{1,1} &
      6.3 & 1.0\\
         & SO & & & & $4.8\times 10^{13}$ & \jj{3}{2}{2}{1} & 8.2 & 2.0\\
         & \htcop        & & & & $8.3\times 10^{11}$ & 1--0 & 9.5 & 0.4\\
     \noalign{\smallskip}
     CL6  & \hdco\ & 50 & 10 & 0.7 & $2.8\times 10^{13}$ & \jj{2}{1,2}{1}{1,1}
     & 4.7 & 4.4\\
        & SO  & & & & $1.3\times 10^{14}$ & \jj{3}{2}{2}{1} & 6.7 & 7.2\\
        & \htcop         & & & & $4.6\times 10^{11}$ & 1--0 & 5.0 & 0.4\\
     \noalign{\smallskip}
     ICLN & \hdco\  & 1000 & 20 & 0.2 & $3.0\times 10^{13}$ &
     \jj{2}{1,2}{1}{1,1} & 3.9 & 3.6\\
        & SO  & & & & $6.6\times 10^{13}$ & \jj{3}{2}{2}{1} & 4.9 & 2.9\\
        & \htcop         & & & & $4.0\times 10^{11}$ & 1--0 & 4.2 & 0.3\\ 
     \noalign{\smallskip}
     ICLS & \hdco\  & 1000 & 20 & 0.2 & $2.5\times 10^{13}$&
     \jj{2}{1,2}{1}{1,1} & 4.0 & 6.1\\  
        & SO  & & & & $7.6\times 10^{13}$ & \jj{3}{2}{2}{1} & 5.0 & 4.9\\
        & \htcop         & & & & $4.0\times 10^{11}$ & 1--0 & 3.8 & 0.5\\

     \noalign{\smallskip}\hline\noalign{\medskip}
   \end{tabular}

   \begin{minipage}{107mm}
     \begin{list}{}{}
       \item[$^a$]Beam-averaged column densities. 
       \item[$^b$]Calculated for the lower frequency transition (the best
         determined in the observations).
     \end{list}
   \end{minipage}
 \end{table*}

\subsubsection{Results of the RADEX analysis: uncertainties}
 \label{results_uncertain}

  Table~\ref{radextable} shows the best-fit results we obtained from our
  $\chi_{\nu}^2$ analysis using RADEX on the set of lines from the \hdco, SO,
  and \htcop\ molecules, for which at least 3 transitions have been detected
  at each position. Table~\ref{table_physparam} shows the best set of
  parameters that we found for \cthd, HCN, and SO$_2$, assuming the
  temperature, volume density and size derived from the previous best fit,
  plus the results from the hyperfine-line fitting for CCH, CN, and NO.

  We estimate that the volume density determination has an uncertainty of
  50--70\% of the best fit value (shown in Table 4) at the 1-sigma level. The
  column densities, $N(X)$, are less well constrained than the volume
  density. The derived column densities of all the molecules have an
  uncertainty of $\sim65$\% on average from the best-fit value for the CL1
  position. For the other positions, the column densities are only constrained
  within a factor of 2 of the best value.

\begin{figure}
   \begin{center}
     \includegraphics[angle=-90,width=\columnwidth]{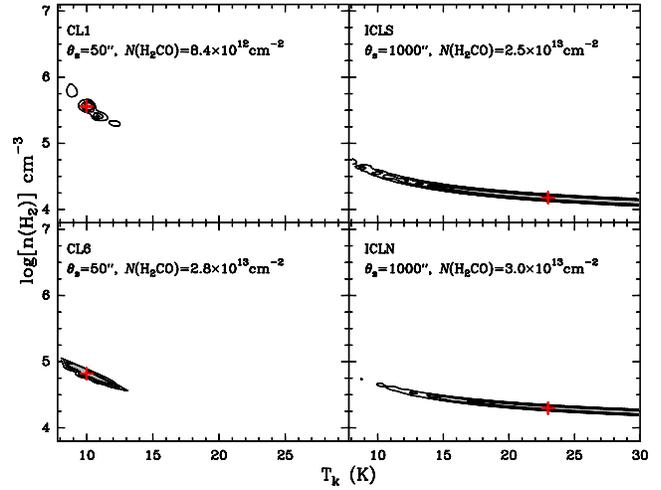}
   \end{center}
 \caption{Contours of constant $\chi_{\nu}^2$ value (the goodness-of-fit) over
   the parameter space spanned by the volume density, $n(\mathrm{H_2}$), and
   $T_\mathrm{K}$ for fixed values of the \hdco\ column density,
   $N(\mathrm{H_2CO}$), for each of the observed positions. The contours mark
   the confidence levels of 68.3\%, 90.0\%, and 99.9\%. Red (grey) crosses
   mark the best-fit parameters for each position.}
 \label{figradex}
 \end{figure}

  We alternatively explored how well constrained were the results of our
  analysis by fixing the \hdco\ column density and the filling-factor for each
  position to the best fit values, and running a range of models with
  different kinetical temperatures and volume densities. Figure~\ref{figradex}
  shows the goodness-of-fit over the parameter space spanned by
  $n(\mathrm{H_2}$) and $T_\mathrm{K}$. Similarly to the results in
  Table~\ref{radextable}, $n(\mathrm{H_2}$) is relatively well-constrained for
  all positions, but, except for CL1, that is not the case for
  $T_\mathrm{K}$. Temperatures are rather badly determined for the two
  inter-clump positions, probably due to the lower volume density and the
  lower SNR of the 1-mm lines. The best fit values of $T_\mathrm{K}$ show
  clear differences between the CL1 and CL6 positions,
  $T_\mathrm{K}\simeq10$~K, and the inter-clump positions,
  $T_\mathrm{K}\simeq23$~K (see also Sect.~\ref{results_ntt}). However, the
  uncertainties in the inter-clump positions are too large to be
  significant. We adopt a value of 20~K for the inter-clump gas, since their
  lower visual extinction may allow the interstellar UV radiation to heat the
  gas \citep[see e.g.,][]{Nejad99}.

  As shown in Fig.~\ref{fobspos}, the 2 and 3 mm beams of the telescope at
  both inter-clump positions could have contribution from higher density
  gas. This effect could be important for the ICLS position, which is the one
  we studied more carefully (the effect of the emission of CL6 on ICLN is
  probably very small because the emission in CL6 is much less intense). We
  used the 1.2-mm continuum map, with an angular resolution $\sim10\farcs5$,
  to estimate the contribution to the intensity of the lines at the ICLS
  position coming from emission from the CL1 position.  In order to have an
  upper limit estimate of the error beam of the telescope at 2 and 3 mm, we
  convolved the 1.2-mm continuum map with a Gaussian with the same HPBW as the
  main beam of the telescope at 2 and 3 mm obtained from \citet{Greve98} and
  \citet{Bensch01}. Then, we compared the intensities at the clump and
  inter-clump positions for the original and the convolved maps. We calculated
  the contribution of the error beam at the CL1 position to the \hdco, SO, and
  \htcop\ lines at the ICLS position as representatives of the different beams
  we used. We found that the error beam contribution to the molecular line
  emission in ICLS can be $\la0.1$--0.5~K, or $\la$18--50\% range of the
  intensity of the detected lines.  Thus, though not negligible, we expect a
  minor contribution of the higher density gas to the lines in the ICLS
  position. In order to check how much the results of the analysis for the
  ICLS position can be affected by this contribution, we remade the analysis
  after subtracting the estimated intensity coming from the CL1 position in
  the lines measured in ICLS. The resulting number density for ICLS becomes
  $\sim3$ times smaller, $\sim6.6\times10^3$\cmt, while the column densities
  of \hdco\ and SO are $\sim2$ times larger, and \htcop\ remains practically
  constant.

\subsubsection{Results of the RADEX analysis: volume density, excitation
  temperature and line opacity}
 \label{results_ntt}

  The line intensities and line ratios of our data are best reproduced if we
  use beam-filling factor and kinetical temperature, $\eta^i_{bf}<1$ and
  $T_\mathrm{K}\simeq10$~K, for CL1 and CL6, but $\eta^i_{bf}\simeq1$ and
  $T_\mathrm{K}\simeq20$~K give better results for the two ``inter-clump''
  positions. We performed an additional test to check if the ``best-values''
  of the filling-factors are consistent with previous data. From the maps by
  MGE05, we calculated the main beam antenna temperatures at each of the four
  positions after convolving the maps with beams of FWHM 14'' and 28'' and
  tried to estimate the size of the ``apparent'' emitting region. The results
  for the ICLN and ICLS positions agree with $\eta^i_{bf}\simeq1$. For CL1 and
  CL6, the results agree with $\theta_s\sim30''$ and $45''$, respectively, not
  far from the adopted value $\theta_s \sim50''$.

  The volume densities yielded by the $\chi_{\nu}^2$ analysis show a clear
  distinction between the physical characteristics of the four positions
  traced by the molecular line emission. The volume density is highest in CL1,
  $\sim3.6\times10^5$\cmt, $\sim5$ times larger than the density derived for
  CL6, and more than one order of magnitude larger than for the inter-clump
  positions.

 \begin{table}
 \caption{Physical parameters for the rest of the molecules observed in the
   four positions of LDN 673 obtained from the best-fit models after fixing
   $T_\mathrm{K}$, $n(\mathrm{H_2})$, and $\theta_s$ to the values reported in
   Table~\ref{radextable},} or from the hyperfine structure fitting (see
 Sect.~\ref{restmolecules}).
 \label{table_physparam}

   \centering
   \begin{tabular}{llcccc}
     \hline\noalign{\smallskip}
     Position & Species &$N(X)\,^a$ & Transition & $\Tex\,^b$ & $\tau^b$\\
     &      & (cm$^{-2}$)        & &  (K)  & \\
     \noalign{\smallskip}\hline\noalign{\smallskip}
     CL1  & \cthd  & $4.5\times10^{12}$  & \jj{2}{1,2}{1}{0,1} & 6.5 & 0.9\\ 
          & HCN    & $1.5\times10^{13}$  & 3--2                & 4.9 & 6.9\\
          & SO$_2$ &  $3.7\times10^{12}$ & \jj{3}{1,3}{2}{0,2} & 4.5 & 0.2 \\
     \noalign{\smallskip}\hline\noalign{\smallskip}
     CL6  & \cthd  & $1.3\times10^{12}$  & \jj{2}{1,2}{1}{0,1} & 3.4 & 0.6\\
     \noalign{\smallskip}\hline\noalign{\smallskip}
     ICLN & \cthd  & $6.0\times10^{11}$  & \jj{2}{1,2}{1}{0,1} & 3.3 & 0.3\\
     \noalign{\smallskip}\hline\noalign{\smallskip}
     ICLS & \cthd  & $1.9\times10^{12}$  & \jj{2}{1,2}{1}{0,1} & 3.2 &  0.9\\
          & SO$_2$ & $5.3\times10^{12}$  & \jj{3}{1,3}{2}{0,2} & 3.2 & 0.6 \\
     \noalign{\smallskip}\hline\noalign{\smallskip}
     \multicolumn{5}{c}{Hyperfine structure calculation}\\
     \noalign{\smallskip}\hline\noalign{\smallskip}
     CL1  & CCH    & $1.7\times10^{13}$  & 1--0                & 5.7 & 0.3\\
          & CN     & $3.4\times10^{13}$  & 1--0                & 4.3 & 5.9\\
          & NO     & $5.3\times10^{14}$  & \jf{3}{2}{5}{2}--\jf{1}{2}{3}{2} &
     5.4 & 0.4\\ 
     \noalign{\smallskip}\hline\noalign{\smallskip}
     CL6  & CCH    & $4.7\times10^{12}$  & 1--0                & 4.5 & 0.1\\
          & CN     & $9.1\times10^{12}$  & 1--0                & 3.3 & 1.4\\
          & NO     & $2.2\times10^{14}$  & \jf{3}{2}{5}{2}--\jf{1}{2}{3}{2} &
     7.1 & $<0.1$\\ 
     \noalign{\smallskip}\hline\noalign{\smallskip}
     ICLN & CCH    & $5.3\times10^{12}$  & 1--0                & 4.1 & 0.1\\ 
          & CN     & $7.3\times10^{12}$  & 1--0                & 3.1 & 1.1\\
          & NO     & $1.7\times10^{14}$  & \jf{3}{2}{5}{2}--\jf{1}{2}{3}{2} &
     5.0 & 0.1$^c$\\ 
     \noalign{\smallskip}\hline\noalign{\smallskip}
     ICLS & CCH    & $7.0\times10^{12}$  & 1--0                & 3.9 & 0.2\\
          & CN     & $1.1\times10^{13}$  & 1--0                & 3.0 & 2.2\\
          & NO     & $2.5\times10^{14}$  & \jf{3}{2}{5}{2}--\jf{1}{2}{3}{2} &
     5.2 & 0.1$^c$\\  
     \noalign{\smallskip}\hline\noalign{\medskip}
   \end{tabular}

   \begin{minipage}{80mm}
     \begin{list}{}{}
       \item[$^a$]Beam-averaged column densities.
       \item[$^b$]Calculated for the lower frequency transition (the best
         determined in the observations).
       \item[$^c$] Calculated after fitting the spectra assuming optically thin
         lines.
     \end{list}
   \end{minipage}
 \end{table}

  Table~\ref{radextable} also shows the values of excitation temperatures,
  $\Tex$, and line opacities, $\tau$, yielded by the ``best-fit'' models for
  each position and molecule: \hdco\ (\jj{2}{1,2}{1}{1,1}), SO
  (\jj{3}{2}{2}{1}), and \htcop\ (1-0). We can see that the \hdco\ and SO
  lines tend to be optically thick or very thick, especially for the CL6,
  ICLN, and ICLS positions, which is also reflected in apparent larger values
  of the column density, compared to the ones found at CL1. This effect is
  possibly due to the lower volume density in CL6, ICLN, and ICLS. The
  rotational levels involved in the \hdco\ and SO lines are not of the
  fundamental rotational level, which makes them more difficult to populate at
  lower densities. In order to reproduce the observed line intensities, we
  need to compensate by using higher column densities (and line
  opacities). This effect is not seen for the \htcop\ lines, involving the
  fundamental rotational level, which are relatively optically thin and with
  CL1 column densities larger or similar to the ones found at the other
  positions.

  The excitation temperatures derived for the lower frequency transition of
  the molecules of Table~\ref{table_physparam} lie in a range $\sim$3--7~K,
  which is very similar to the values of $\Tex$ previously derived for CS
  (2--1) and \ndh\ (1--0) (MGE03, MGE05). Most of the lines are also
  relatively optically thin, with a few exceptions.

\subsubsection{Results of the RADEX analysis: column density}

  \hdco\ and SO column densities are lowest at CL1, by factors of
  $\sim3.0$--3.6 and $\sim1.4$--2.8, respectively, compared to the other
  positions, while $N$(\htcop) is the largest at CL1 by a factor of
  $\sim2$. Both ICLN and ICLS show rather similar values for the column
  densities of the three molecules.

  The column density variations for CCH, \cthd\, CN, and NO follow similar
  trends, with small differences. The CL1 position shows the largest column
  densities for all these molecules, while ICLS has the second largest column
  density. Column densities in CL6 tend to be slightly larger or similar to
  the ones in ICLN.  The column density of SO$_2$, which was only detected in
  two positions, is also $\sim50$\% larger for ICLS than for CL1.

\subsection{Dust continuum emission}
 \label{dustcont}

  We calculated the mass of the molecular gas traced by the dust from the
  continuum emission flux density \citep{Frau10}. Table~\ref{tdustcond} gives
  the H$_2$ mass, and H$_2$ averaged column and volume density contained
  inside the FWHM contour of each of the four condensations found in the
  1.2-mm continuum map, assuming a dust temperature of 10~K.

  The calculated masses of the four condensations range from 1.6~\mo\ for the
  SE condensation, which is also the smallest one ($\sim 0.05$~pc), to
  5.1~\mo\ for the CL1/SMM5 condensation. The H$_2$ volume densities for the
  three centrally peaked condensations are rather similar, $\sim3\times
  10^5$\cmt, which is very close to the value derived for the CL1 position
  from the spectral line observations. The weaker and more diffuse emission of
  the SMM6 condensation is reflected in a volume density $\sim4$ times lower,
  $7\times10^4$\cmt, slightly larger than the value derived for the CL6
  position. Interestingly, the condensations associated with CL1/SMM5 and SMM3
  have similar extensions ($\sim 0.07$ vs $\sim 0.06$ pc), volume densities,
  and, to a lesser extent, column densities.

  In order to estimate the abundances of the observed molecular species (see
  Sect.~\ref{abundances}), we also calculated the H$_2$ column density traced
  by the dust continuum emission at the four spectral positions using three
  different angular resolutions: $27''$, $22''$ and $17''$, corresponding to
  the angular resolution of the spectral line observations at 3.4, 2.7 and
  2.1~mm, respectively (Table~\ref{tdustmass}). We adopted a dust temperature
  value of 10~K for the four positions, given that because of the lack of any
  internal heating source and that the densities of the gas seem to be
  $\ga10^4$\cmt\ we do not expect higher dust temperatures
  \citep{Goldsmith01}.  The column densities at the CL1 position are
  significantly larger than at the other three positions, by factors of $\sim
  3$--4 with respect to CL6 and ICLN, and of $\sim$5--10 with respect to ICLS.

 \begin{table}
 \caption{H$_2$ average column density traced by the dust continuum emission
   on the positions of the spectral line observations for three different beam
   sizes.}  
 \label{tdustmass}
   \centering
   \begin{tabular}{lcrrr}
     \hline\noalign{\smallskip}
     && \multicolumn{3}{c}{$N(\mathrm{H_2})\,^a$}\\
     & $\tdust$ & \multicolumn{3}{c}{$(10^{21}$\cmd)}\\
     \noalign{\smallskip}\cline{3-5}\noalign{\smallskip}
     Position & (K) & $27''$ & $22''$ & $17''$ \\
     \noalign{\smallskip}\hline\noalign{\smallskip}
     CL1	   & 10 & 31.1  & 34.8  & 40.1 \\
     CL6	   & 10 & 10.8  & 10.9  & 10.2 \\
     ICLN	   & 10 &  9.0  &  9.5  & 11.2 \\
     ICLS	   & 10 &  6.0  &  5.5  &  4.3 \\
     \noalign{\smallskip}\hline\noalign{\medskip}
   \end{tabular}

   \begin{minipage}{55mm}
     \begin{list}{}{}
       \item[$^a$] Using the parameters given in note $^c$ of
         Table~\ref{tdustcond}.
     \end{list}
   \end{minipage}
 \end{table}

\subsubsection{Comparison of dust and CS emission}
 \label{sect_dustcs}

\begin{figure}
   \begin{center}
     \includegraphics[width=75mm]{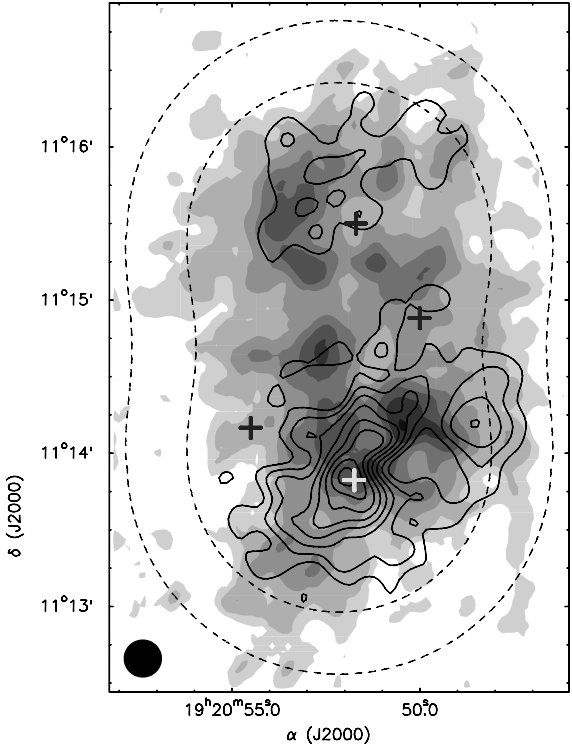}
   \end{center}
 \caption{Greyscale of the CS (2-1) integrated emission (from MGE05)
   overlapped with the contour map of the 1.2 mm dust emission. The angular
   resolution of both images is $\simeq 15''$. The contours of the dust
   continuum map start at 14\mjy with increments of 7\mjy; the contours of the
   CS (2--1) map begin at 0.496~K with increments of 0.248~K. The two maps
   were convolved with the primary beam response of the BIMA two-point mosaic
   observations presented by MGE05. The inner and outer dashed lines show the
   0.5 and 0.25 level of the BIMA primary beam response. The thick crosses
   show the four positions observed with the IRAM 30m telescope.}
 \label{fcsgrey}
 \end{figure}

  We convolved the dust map with the 2-point mosaic response of the BIMA
  telescope in order to properly compare the CS (2--1) and 1.2~mm dust maps.
  Figure~\ref{fcsgrey} shows the overlap of the CS (2--1) integrated emission
  from MGE05 with the convolved dust continuum emission at the same angular
  resolution, $15''$. In general the CS and 1.2~mm emissions are not
  correlated.  For the CL1/SMM5 condensation, the CS emission peak is slightly
  displaced with respect to the dust emission peak, by $\simeq 2''$. The
  western condensation, SMM3, lies just at the limit of the BIMA primary beam,
  but it seems to be associated with the CS clump 12 (following MGE05
  notation). The other CS clumps found through the CLUMPFIND analysis done by
  MGE05 do not show any clear dust emission peaks. Some of these clumps in the
  northern part of the map could be associated with the diffuse dust emission
  arising from SMM6. More interesting is the existence of CS emission in the
  central parts of the map with very weak or no dust emission.

\begin{figure}
    \begin{center}
     \includegraphics[width=75mm]{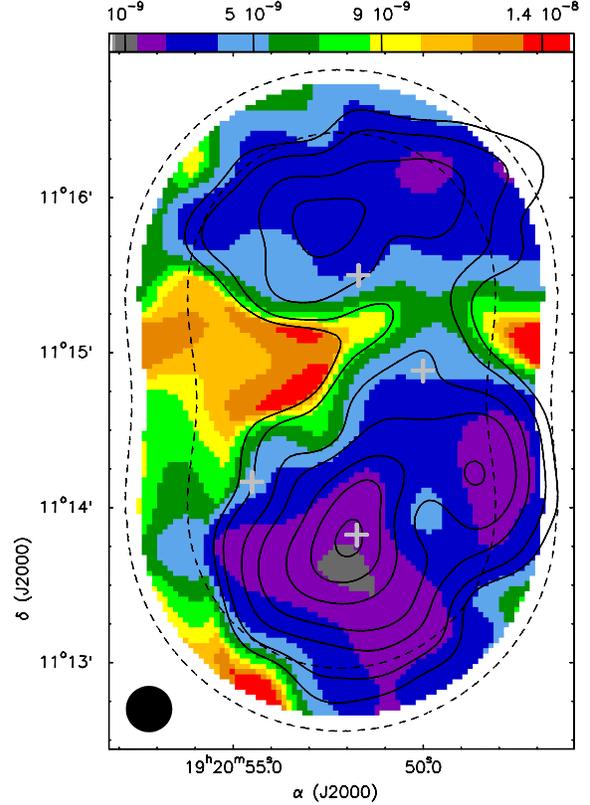}
   \end{center}
 \caption{Colour image of the map of the CS abundance, overlapped with the
   contour map of the 1.2 mm continuum emission. The angular resolution of
   both images is $25''$. The two maps were convolved with the primary beam
   response of the BIMA two-point mosaic observations presented by MGE05. The
   inner and outer dashed lines show the 0.5 and 0.25 level of the BIMA
   primary beam response.  The contours are 10\%, 15\%, 25\%, 35\%, 55\%,
   75\%, and 95\% of the 1.2 mm continuum peak emission, (66.2\mjy). The grey
   crosses show the four positions observed with the IRAM 30m telescope.}
 \label{fdusttocs}
 \end{figure}

  To estimate the CS molecular abundance in the observed BIMA field (see
  Fig.~\ref{fcsgrey}) we first convolved the CS and dust continuum maps with a
  beam size of 25$''$, in order to measure the dust emission with a higher SNR
  in the regions where it is diffuse and weak.  We then calculated the CS
  column density assuming LTE conditions, an excitation temperature of
  $\Tex$=5~K, and a CS (2--1) optical depth line of 3, values derived by MGE05
  from multitransitional analysis of the CS and C$^{34}$S molecules. The H$_2$
  column densities were derived from the dust map adopting a temperature
  $\tdust$=10~K and dust absorption coefficient of 0.5~cm$^2$~g$^{-1}$.
  Figure~\ref{fdusttocs} shows the resulting CS abundance overlapped with the
  dust emission at the same $25''$ angular resolution. The CS abundance
  changes about one order of magnitude, from $X$(CS)$\sim 10^{-9}$ around the
  CL1/SMM5 clump to $X$(CS)$\sim 10^{-8}$ about $1'$ north of CL1/SMM5, closer
  to the position of the CL6 CS clump. CS abundances tend to be lower at the
  positions of the detected dust condensations.

\subsection{Molecular abundances}
 \label{abundances}

\begin{figure}
   \center
   \includegraphics[width=8cm]{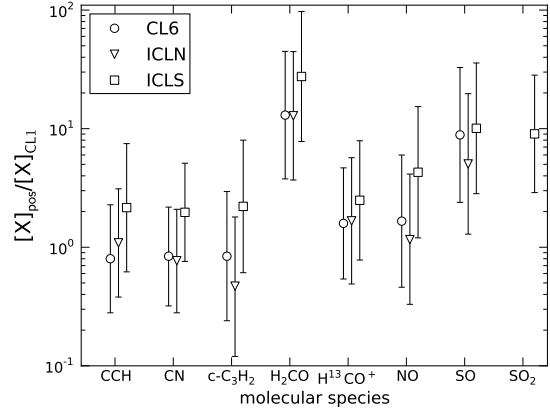}
 \caption{Relative abundances of the molecules detected at the CL6
   \textit{(circles)}, ICLN \textit{(inverted triangles)}, and ICLS
   \textit{(squares)} positions with respect to the abundance at CL1. The
   error bars indicate the estimated uncertainties in the determination of the
   molecular abundances.}
 \label{fig_abundances}
 \end{figure}

  For each position, we calculated the molecular abundances of our molecules
  (Table~\ref{tabundances}) from the ratio of the beam-averaged molecular
  column density to the corresponding beam-averaged H$_2$ column density from
  Table~\ref{tdustmass}. Taking into account the uncertainties in the
  calculation of both sets of column densities, we estimate that the
  uncertainties in the determination of the molecular abundances should be
  between factors of 2--2.5. For easier comparison, we plotted in
  Fig.~\ref{fig_abundances} the upper and lower limits that we estimated for
  the abundances of each molecular species in CL6, ICLN, and ICLS relative to
  the abundance in CL1.

  The abundances of CCH, CN, and \cthd\ at CL1 are similar to the values
  of the other positions, but for the rest of the molecules the abundances are
  clearly lower, by a factor of $\sim2$ for \htcop\ and NO, and of $\sim10$
  for SO, SO$_2$, and \hdco. On the other hand, the abundances are
  systematically larger for all species at ICLS, but only by a factor of
  $\la3$ with respect to the abundances at CL6 and ICLN.

 \begin{table}
 \caption{Molecular abundances calculated from the ratio of the beam-averaged
   molecular column density to the beam-averaged H$_2$ column density derived
   from the dust continuum emission.}
 \label{tabundances}
 
  \centering

   \begin{tabular}{llcl}
     \hline\noalign{\smallskip}
     Species & Position &  Beam$\,^a$ & \multicolumn{1}{c}{[X]/[H$_2$]} \\
     \noalign{\smallskip}\hline\noalign{\smallskip}
     CCH     & CL1      & $27''$ & $5.4\times10^{-10}$ \\
             & CL6      &        & $4.3\times10^{-10}$ \\
             & ICLN     &        & $5.8\times10^{-10}$ \\
             & ICLS     &        & $1.2\times10^{-9}$ \\
     \noalign{\smallskip}
     CN      & CL1      & $22''$ & $9.9\times10^{-10}$ \\
             & CL6      &        & $8.3\times10^{-10}$ \\
             & ICLN     &        & $7.6\times10^{-10}$ \\
             & ICLS     &        & $2.0\times10^{-9}$ \\
     \noalign{\smallskip}
     \cthd   & CL1      & $27''$ & $1.4\times10^{-10}$ \\
             & CL6      &        & $1.2\times10^{-10}$ \\
             & ICLN     &        & $6.7\times10^{-11}$ \\
             & ICLS     &        & $3.2\times10^{-10}$ \\
     \noalign{\smallskip}
      HCN    & CL1      & $12''$ & $3.0\times10^{-10}$ \\
     \noalign{\smallskip}
     \hdco   & CL1      & $17''$ & $2.1\times10^{-10}$ \\
             & CL6      &        & $2.7\times10^{-9}$ \\
             & ICLN     &        & $2.7\times10^{-9}$ \\
             & ICLS     &        & $5.8\times10^{-9}$ \\
     \noalign{\smallskip}
     \htcop   & CL1      & $27''$ & $2.7\times10^{-11}$ \\
             & CL6      &        & $4.2\times10^{-11}$ \\
             & ICLN     &        & $4.4\times10^{-11}$ \\
             & ICLS     &        & $6.6\times10^{-11}$ \\
     \noalign{\smallskip}
     NO      & CL1      & $17''$ & $1.3\times10^{-8}$ \\
             & CL6      &        & $2.2\times10^{-8}$ \\
             & ICLN     &        & $1.5\times10^{-8}$ \\
             & ICLS     &        & $5.7\times10^{-8}$ \\
     \noalign{\smallskip}
     SO      & CL1      & $22''$ & $1.4\times10^{-9}$ \\
             & CL6      &        & $1.2\times10^{-8}$ \\
             & ICLN     &        & $6.9\times10^{-9}$ \\
             & ICLS     &        & $1.4\times10^{-8}$ \\
     \noalign{\smallskip}
     SO$_2$  & CL1      & $22''$ & $1.1\times10^{-10}$ \\
             & ICLS     &        & $9.6\times10^{-10}$ \\
     \noalign{\smallskip}\hline\noalign{\medskip}
   \end{tabular}

   \begin{minipage}{6cm}
     \begin{list}{}{}
       \item[$^a$] Beam-size over which the molecular and H$_2$ column
         densities were averaged.
     \end{list}
   \end{minipage}
 \end{table}

\section{Discussion}
 \label{discussion}

\subsection{Fragmentation in the core}

  The most intense emission in the central region of the 1.2-mm dust continuum
  map is contained inside the HPBW of the low-angular resolution \nht\ map of
  \citet{Sepulveda11}. The BIMA observations of MGE03 also only detected \ndh\
  emission in a region almost coinciding with CL1/SMM5 (the other dust
  continuum sources lie outside the BIMA primary beam). The dust and \ndh\
  emissions in CL1/SMM5 are similarly elongated in an approximately N--S
  direction with a close proximity between both emission peaks. A second
  clump, located just North of CL1/SMM5 could also be present in the dust
  emission, but it is not so clearly seen as in the \ndh\ map.  The sizes of
  the CL1/SMM5 condensation in the continuum and molecular (CS and \ndh) maps
  are very similar, $\sim$0.07--0.09~pc, but the gas mass traced by the dust
  is $\sim$ 4 times larger than the masses estimated, assuming standard
  abundances, from the CS and \ndh\ maps by MGE05.

  We showed in Sect.~\ref{sect_dustcs} that the CS and dust emissions do not
  seem to be particularly correlated (see Figs.~\ref{fcsgrey} and
  \ref{fdusttocs}): the most intense CS emission around CL1/SMM5 follows the
  edges of the most intense dust emission and many of the clumps detected in
  the combined FCRAO and BIMA maps are located in regions with no or very weak
  dust emission. These results suggest that there is some low
  density/small-size structure which is revealed by the CS lines, but it is
  not detected at 1.2-mm, probably because the sensitivity of our observations
  is not high enough or due to the spatial filtering of the chopping bolometer
  observations.

  The SMM8 dust condensation is also coincident with relatively weaker \htcop\
  and CS emission detected with the FCRAO telescope by MGE05 at the tip of a
  very weak NW-SE filamentary structure or thin ridge found in the molecular
  line maps. This filamentary structure is undetected in the dust continuum
  map.

  In order to study the fragmentation in the core, we calculated the
  Jeans-length corresponding to the densities of the two main dust
  condensations, CL1/SMM5 and SMM3, following \citet{Hartmann98book}:
  \begin{equation}
    \lambda_\mathrm{J} = \left ( \frac{\pi a^2}{G\rho} \right)^{1/2} =
    0.19~\mathrm{pc} \left (\frac{T}{10 \mathrm{K}} \right )^{1/2} \left (
    \frac{n_\mathrm{H_2}}{10^4 \mathrm{cm^{-3}}} \right )^{-1/2} 
  \end{equation}
  with the sound speed $a$, the gravitational constant $G$, and the density
  $\rho$. The separation between the condensations is $\sim50''$, which at the
  distance of LDN~673 corresponds to $\sim0.073$ pc ($\sim15000$ AU). We used
  the average of the densities estimated for both condensations from the
  continuum observations, $\sim3.3\times10^5$\cmt, as the volume density,
  $n_\mathrm{H_2}$, and assumed $T=10$~K. We obtained a Jeans-length value of
  $\sim0.033$ pc, about 2 times smaller than the measured separation on the
  sky, and similar to the values found by \citet{Tsitali10} for some other
  condensations in the region. This value of the Jeans-length is also smaller
  than the sizes of both condensations, which is an indication that they may
  still be subject to more thermal fragmentation. Indeed, there is evidence of
  fragmentation in the CL1/SMM5 condensation, since this clump is associated
  with two \ndh\ condensations separated by $\sim30''$ (MGE05), which
  corresponds to $\sim0.044$ pc ($\sim$9000 AU), very close to the
  Jeans-length previously estimated. On the other hand, most of the CS clumps
  detected by MGE05 and not traced by the 1.2-mm dust map have typical sizes
  of 0.04 pc and densities of a few times $10^4$\cmt. For a density of $7
  \times10^4$\cmt\ (the CL6 density), the Jeans-length is 0.07 pc. Thus, these
  clumps are unlikely to have been formed as a result of thermal fragmentation
  and we would need to invoke non-thermal processes (turbulence, MHD waves,
  etc) to explain their existence \citep{FH02, Vazquez05, Klessen05,
  VanLoo08}. This would agree with the transient nature of this clumps
  postulated by MGE05.

\subsection{The nature of the clumps and the inter-clump medium}

  The interpretation of single-point observations of molecular lines as the
  ones presented here have to be taken with caution, because they will never
  provide the same level of detail of the chemistry as a full map
  would. However, our choice of positions with presumed different physical
  conditions, coupled with the dust continuum map, provides us with a sample
  of molecular abundances that can show some of the properties of our core.

  MGE05 already showed that CS was probably depleted in CL1, while \ndh\ was
  clearly centrally condensed. The estimation of the CS abundance in
  Fig.~\ref{fdusttocs} confirms this result for the SMM5, SMM3, and SMM6 dust
  condensations. Additionally, Table~\ref{tabundances} shows that \hdco, SO,
  SO$_2$, and, to a lesser extent, \htcop\ are also depleted for CL1 with
  respect to the abundances at the other positions. On the other hand, the
  abundances of CCH, \cthd, and CN in CL1 do not seem to differ much from the
  values in CL6 and ICLN. Finally, the NO abundance at CL1 is also $\sim40$\%
  lower than the one at CL6, and similarly to what \citet{Akyilmaz07} found in
  L1544, the NO molecule could be partially depleted in CL1, while \ndh\ is
  not.

  The original models of \citet{Taylor96,Taylor98} and
  \citet{Garrod05,Garrod06} required the action of depletion to obtain the
  distribution of molecular abundances that would explain the results of MGE03
  and MGE05. As has been extensively found in different starless cores,
  'early-time', carbon-bearing molecules, would be depleted in the more
  central regions of evolved starless cores, wherein 'late-time',
  nitrogen-bearing molecules, would still be present in the gas \citep[see
    e.~g.,][]{Caselli99,HilyBlant10,Tafalla06}. Thus, the results for CS,
  \ndh, \hdco, SO, and \htcop\ are approximately as expected, given the
  density of the CL1/SMM5 condensation. However, the low or no depletion of
  CCH, \cthd, and CN is much more unexpected.  \citet{Frau11} also found a
  similar result in several cores of the Pipe Nebula, where SO shows
  depletion, but \cthd\ and CCH are more intense in the cores with higher
  A$_\textrm{V}$.

  CL6 and the inter-clump positions show a poorer variety of molecules
  compared to that of CL1. Additionally, there is little evi\-dence of
  depletion of the detected molecules -- the abundances of early-time
  molecules are usually the largest in the inter-clump positions -- with no
  detection of \ndh. All this seems to indicate that the gas in CL6, ICLN, and
  ICLS is chemically young. The gas density in CL6 is more similar to the
  densities found in the inter-clump medium (see Sect.~\ref{results_ntt}) than
  to the density in CL1. However, the number density derived for CL6 is still
  a factor of $\sim3.5$ larger with respect to the two inter-clump positions,
  which indicates that the physical conditions are probably different. This
  factor is significant and probably a lower limit to the true density
  contrast, because the probable contribution to the emission from nearby
  clumps will have the effect of measuring a density higher than the true
  inter-clump density in ICLN and ICLS (see Sect~\ref{results_uncertain}). In
  any case, it seems clear that the inter-clump positions are denser than
  initially expected, probably from a few times $10^3$ to $10^4$\cmt.

\subsubsection{Comparison with models of transient clumps}

  In order to have a more quantitative comparison, we compared the results of
  our observations to the ``standard'' model of \citet{Garrod05,Garrod06}. We
  found that this model is able to explain reasonably the chemistry of the CL6
  and ICLN positions. However, we used a ``high-density'' model that reached a
  higher peak density, $\sim5.5\times10^5$\cmt, in $10^6$ yr
  \citep{Garrod05thesis} to match better the chemistry in CL1. We did not
  compare the models to ICLS due to the contamination issues discussed in
  Sect.~\ref{results_uncertain}. We consider that the real evolutionary
  properties of the gas we observed are pro\-ba\-bly between these two models
  and some scaling of abundances would be needed to really fit our results,
  which is outside the purpose of this discussion. Given the traditional
  difficulty of matching the values of chemical models to observations, we
  looked for the time ranges in which the abundances of most of the molecules
  in the models ($>70\%$) were within a factor of 3 from the observed values,
  in order to account for the multiple sources of uncertainties in both
  observations and models.

  The molecular abundances in CL1 are compatible with a clump in the
  ``high-density'' model in a time range of 3--7$\times10^5$ yr. \hcop\ is the
  main outlier, depending on the $^{12}$C/$^{13}$C abundance ratio used, and
  ``late-time'' molecules, NO and SO, tend to push the time range to later
  times. The abundances in CL6 are better reproduced with a clump following
  the standard model at times either 5--9$\times10^5$ yr or
  1.4--$1.6\times10^6$ yr. So, the clump in CL6 could be very close to the
  peak density maximum or just past it in the dissipation phase. We also found
  that the results for ICLN could be compatible with a clump of
  0.9--1.6$\times10^6$ yr for the standard model. This could explain the
  relatively low contrast in density between CL6 and ICLN, if we think that
  the gas in ICLN could be the remains of a clump that has almost dispersed
  and what we found was lower density gas enriched by the molecular abundances
  of previous stages. This would reinforce one of the predictions of
  \citet{Garrod06} models: that the formation and dispersion of transient
  clumps would result in a chemistry of the gas generally ``young'', except in
  the denser and more evolved clumps, and a gas chemical composition that
  would show signs of ``enrichment'' of abundances of some molecular species
  in the lower density material.

  \citet{Garrod06} also showed that a distribution of transient clumps at
  different stages of evolution (different points in time) could qualitatively
  reproduce the medium- and high-angular resolution maps of MGE03 and
  MGE05. Our results here suggest that those clumps could not only be in
  different stages of evolution but also reach different peak densities, i.e.\
  follow different evolutionary patterns. The mix of clumps with different
  evolutionary properties can shed a different light on the discussion about
  the mole\-cular depletion between different clumps. We found that, for the
  times best fitted by the models, the differences in the abundances of CCH,
  \cthd, and CN between the ``high-density'' and ``standard'' models are
  considerably smaller than for later-time molecules, such as NO. This would
  simply explain the apparent low or no depletion of the former molecules
  compared to the latter.

\section{Conclusions}

  We studied the properties of the clump and inter-clump gas in the CS
  starless core of LDN~673 using the 30-m IRAM telescope through the emission
  of several spectral lines in the 3-, 2-, and 1-mm bands, in four positions
  associated with already detected, but physically and chemically different,
  clump gas (positions CL1 and CL6), or with inter-clump gas (positions ICLN
  and ICLS). We detected 19 spectral transitions of 10 molecular species at
  least in one of the four positions. We complemented the spectral
  observations with the mapping of the 1.2-mm dust continuum emission in
  LDN~673, which allowed us to obtain a more reliable estimate of the volume
  density of the gas and of the abundances of the detected molecules. The main
  results of our study are:

  \begin{enumerate}

    \item The dust continuum observations revealed four emission condensations
      in the region, roughly coinciding with SMM sources found by
      \citet{Visser02}, three of them with a round shape and centrally peaked
      emission: CL1/SMM5 is the most intense and coincides with the \ndh\
      clump found by MGE03; SMM3, located $\sim50''$ to the West, was not
      previously found in the molecular observations; and SMM8, located
      $\sim2\farcm5$ SE of the centre of the map, and coinciding with an
      emission enhancement in the \cts\ and \htcop\ maps of MGE05. Finally,
      there is a more diffuse emission strip extended E--W $\sim2'$ north of
      CL1/SMM5. The masses of these condensations range from 1.6 to 5.1~\mo\
      and sizes 0.05--0.07~pc. The northern condensation is sensibly more
      diffuse.

    \item We made a radiative transfer analysis using the RADEX code to
      determine the molecular column densities, volume density, and kinetical
      temperature at each position. The best fits to our observations found
      $T_\mathrm{K}\simeq10$~K for the CL1 and CL6, and $\simeq20$~K for the
      ICLN and ICLS positions; and volume densities ranging from
      $\sim3.6\times10^5$\cmt\ at CL1 to $\sim 1.7\times10^4$\cmt\ at ICLS.

    \item CL1 presents the largest column densities for most of the molecules
      (\htcop, \cthd, CCH, CN, and NO), but the smaller column densities for
      \hdco\ and SO. The estimated molecular abundances of most of the
      molecules, are smallest at CL1, while the abundances at CL6 and ICLN
      tend to be very similar.

    \item The comparison of the dust continuum emission to previous
      interferometer and single-dish observations shows very little or no
      correlation between the CS (2--1) and the 1.2-mm emission, with most of
      the clumps found in the CS emission not detected in the continuum
      observations. On the other hand, the dust continuum and \ndh\ emissions
      are much more similar.

    \item We found that the condensations detected in the dust continuum map
      may be subject to more thermal fragmentation, which might be already
      happening in CL1/SMM5. However, most of the clumps detected in CS, but
      undetected in the dust continuum observations, are unlikely to have been
      formed by thermal fragmentation, and would need other mechanisms to
      explain their formation. These clumps were proposed to be transient by
      MGE05.

    \item The chemistry of CL1 appears to be much more evolved than for the
      other positions, with signs of depletion of several molecular species
      (CS, \hdco, SO, \htcop, and partially NO), but also relatively
      unexpected high abundances of CCH, \cthd, and CN. The density contrast
      between CL6 and the two inter-clump positions, which are denser that
      initially expected, is relatively low. The gas at the CL6 and
      inter-clump positions seems to be generally chemically ``young''.

  \end{enumerate}

  In summary, the central condensation, CL1/SMM5, is probably approaching the
  'peak-time' state of the models of \citet{Garrod05,Garrod06} and it is also
  the place with a larger probability of future undergoing of star formation
  (see MGE05). The SMM3 condensation is probably in a very similar state, from
  its shape and gas density, but we cannot firmly determine it until further
  observations provide us with more information about its chemistry. The low
  density contrast between CL6 and the inter-clump positions and their similar
  young chemical age seems to support the idea of the presence of lower
  density transient clumps in the core as proposed by \citet{Garrod06}.

\section{acknowledgements}
  O.M.\ is supported by the NSC (Taiwan) ALMA-T grant to the Institute of
  Astronomy \& Astrophysics, Academia Sinica. O.M.\ acknowledges support from
  the National Science Foun\-dation (US) while a postdoc at Ohio State
  University working with Eric Herbst. J.M.G.\ and R.E.\ are supported by
  MICINN grant AYA2008-06189-C03 (Spain). O.M., J.M.G., and R.E.\ are also
  supported by AGAUR grant 2009SGR1172 (Catalonia).


\appendix

\section{Line parameters of the observed lines}

  Tables~\ref{tline_cl1} to \ref{tline_icls} show the line parameters
  obtained using a Gaussian fit for the transitions detected in the CL1, CL6,
  ICLN, and ICLS positions, respectively.  Each table lists the molecular
  species and transition and the four parameters resulting from the Gaussian
  fit: line intensity (in main-beam units), central velocity of the line,
  line-width, and intensity integrated under the Gaussian.

  Table~\ref{tline_hfs} shows the hyperfine structure fit parameters, obtained
  using the HFS method of the CLASS package, for the CN, CCH and NO
  transitions detected at 3-mm. The table lists: position, molecular species
  and transition, A$\tau$, central velocity of the hyperfine line of reference,
  and optical depth of this line.

  Finally, Table~\ref{tline_nd} gives the $3\sigma$ upper limits for the line
  intensities of the transitions not detected at each position.

 \begin{table*}
 \caption{Line parameters for the transitions detected in Clump 1 (CL1)}
 \label{tline_cl1}
   \centering
   \begin{tabular}{lcccc}
     \hline\hline\noalign{\smallskip}
	          &$T_\mathrm{MB}$ &$V_\mathrm{LSR}$& $\Delta V$   & Line area
	         \\ 
     Transition   & (K)            &(km~s$^{-1}$) & (km~s$^{-1}$) &(K
	         km~s$^{-1}$)\\ 
     \noalign{\smallskip}\hline\noalign{\smallskip}
     CCH 2--1     & $0.35\pm0.07$ & $+7.10\pm0.04$ & $0.48\pm0.09$ &
 	        $0.18\pm0.03$\\ 
     \noalign{\smallskip}
     CN 2--1	 & $0.79\pm0.08$ & $+7.00\pm0.01$ & $0.41\pm0.03$ &
 	        $0.35\pm0.02$\\ 
     \noalign{\smallskip}
     \cthd\ \jj{2}{0,1}{1}{1,2} & $1.46\pm0.02$ & $+7.16\pm0.01$ &
 	        $0.49\pm0.01$ & $0.76\pm0.01$\\ 
     \cthd\ \jj{4}{1,4}{3}{0,3} & $0.52\pm0.06$ & $+7.15\pm0.01$ &
 	        $0.51\pm0.03$ & $0.28\pm0.02$\\ 
     \noalign{\smallskip}
     HCN 3--2     & $0.88\pm0.26$ & $+6.84\pm0.03$ & $0.28\pm0.08$ &
 	        $0.27\pm0.06$\\ 
     \noalign{\smallskip}
     \hdco\ \jj{2}{1,2}{1}{1,1} & $1.73\pm0.06$ & $+7.02\pm0.01$ &
 	        $0.54\pm0.02$ & $1.00\pm0.02$\\ 
     \hdco\ \jj{3}{1,3}{2}{1,2} & $1.14\pm0.07$ & $+7.00\pm0.01$ &
                $0.46\pm0.02$ & $0.56\pm0.03$\\ 
     \hdco\ \jj{3}{1,2}{2}{1,1} & $0.75\pm0.06$ & $+7.03\pm0.01$ &
                $0.41\pm0.03$ & $0.33\pm0.02$ \\ 
     \noalign{\smallskip}
     \htcop\ 1--0 & $1.44\pm0.04$ & $+7.23\pm0.01$ & $0.52\pm0.01$ &
 	        $0.79\pm0.02$\\ 
     \htcop\ 2--1 & $1.04\pm0.15$ & $+7.19\pm0.02$ & $0.38\pm0.06$ &
 	        $0.42\pm0.05$\\ 
     \htcop\ 3--2 & $0.39\pm0.13$ & $+7.25\pm0.03$ & $0.23\pm0.09$ &
 	        $0.10\pm0.03$\\ 
     \noalign{\smallskip}
     NO \jf{5}{2}{7}{2}--\jf{3}{2}{5}{2} & $0.39\pm0.12$ & $+7.14\pm0.06$ &
 	        $0.50\pm0.11$ & $0.21\pm0.05$\\ 
     \noalign{\smallskip}
     SO \jj{3}{2}{2}{1} & $3.45\pm0.03$ & $+7.12\pm0.01$ & $0.52\pm0.01$ &
 	        $1.89\pm0.01$\\ 
     SO \jj{4}{5}{3}{4} & $0.17\pm0.04$ & $+7.07\pm0.03$ & $0.27\pm0.06$ &
 	        $0.05\pm0.01$\\ 
     SO \jj{6}{5}{5}{4} & $0.75\pm0.09$ & $+7.03\pm0.01$ & $0.35\pm0.02$ &
 	        $0.28\pm0.02$\\
     \noalign{\smallskip}
     SO$_2$  \jj{3}{1,3}{2}{0,2} & $0.19\pm0.03$ & $+7.03\pm0.03$ &
 	        $0.42\pm0.06$ & $0.08\pm0.01$\\
     \noalign{\smallskip}\hline
   \end{tabular}
 \end{table*}

 \begin{table*}
 \caption{Line parameters for the transitions detected in Clump 6 (CL6)}
 \label{tline_cl6}
   \centering
   \begin{tabular}{lcccc}
     \hline\hline\noalign{\smallskip}
 	         & $T_\mathrm{MB}$&$V_\mathrm{LSR}$& $\Delta V$    & Line area
 	         \\ 
     Transition   & (K)           & (km~s$^{-1}$)  & (km~s$^{-1}$)&(K
 	     km~s$^{-1}$)\\
     \noalign{\smallskip}\hline\noalign{\smallskip}
     \cthd\ \jj{2}{0,1}{1}{1,2} & $0.20\pm0.01$ & $+7.23\pm0.01$ &
 	         $0.46\pm0.03$ & $0.10\pm0.01$\\
     \noalign{\smallskip}
     \hdco\ \jj{2}{1,2}{1}{1,1} & $1.25\pm0.07$ & $+7.16\pm0.01$ &
 	         $0.67\pm0.02$ & $0.90\pm0.03$\\ 
     \hdco\ \jj{3}{1,3}{2}{1,2} & $0.77\pm0.06$ & $+7.20\pm0.01$ &
                 $0.56\pm0.04$ & $0.45\pm0.02$\\ 
     \hdco\ \jj{3}{1,2}{2}{1,1} & $0.38\pm0.05$ & $+7.17\pm0.02$ &
                 $0.48\pm0.06$ & $0.19\pm0.02$\\ 
     \noalign{\smallskip}
     \htcop\ 1--0 & $0.51\pm0.03$ & $+7.35\pm0.01$ & $0.60\pm0.03$ &
 	         $0.32\pm0.01$\\
     \htcop\ 3--2 & $0.35\pm0.10$ & $+7.35\pm0.43$ & $0.26\pm0.12$ &
 	         $0.10\pm0.04$\\
     \noalign{\smallskip}
     SO \jj{3}{2}{2}{1} & $2.87\pm0.04$ & $+7.24\pm0.01$ & $0.52\pm0.01$ &
 	         $1.60\pm0.02$\\
     \noalign{\smallskip}\hline
   \end{tabular}
 \end{table*}

 \begin{table*}
 \caption{Line parameters for the transitions detected in Inter-clump North
   (ICLN)}
 \label{tline_icln}
   \centering
   \begin{tabular}{lcccc}
     \hline\hline\noalign{\smallskip}
	     &$T_\mathrm{MB}$&$V_\mathrm{LSR}$& $\Delta V$    & Line area    \\
     Transition   & (K)     & (km~s$^{-1}$)  & (km~s$^{-1}$)&(K km~s$^{-1}$)\\
     \noalign{\smallskip}\hline\noalign{\smallskip}
     \cthd\ \jj{2}{0,1}{1}{1,2} & $0.10\pm0.02$ & $+7.27\pm0.02$ &
	     $0.40\pm0.05$ & $0.04\pm0.01$\\ 
     \noalign{\smallskip}
     \hdco\ \jj{2}{1,2}{1}{1,1} & $0.79\pm0.07$ & $+7.19\pm0.02$ &
	     $0.95\pm0.04$ & $0.80\pm0.03$\\ 
     \hdco\ \jj{3}{1,3}{2}{1,2} & $0.37\pm0.06$ & $+7.32\pm0.03$ &
             $0.79\pm0.08$ & $0.31\pm0.03$\\ 
     \hdco\ \jj{3}{1,2}{2}{1,1} & $0.14\pm0.04$ & $+7.34\pm0.06$ &
             $0.84\pm0.14$ & $0.12\pm0.02$\\  
     \noalign{\smallskip}
     \htcop\ 1--0 & $0.35\pm0.02$ & $+7.34\pm0.01$ & $0.61\pm0.03$ &
	     $0.23\pm0.01$\\ 
     \noalign{\smallskip}
     SO \jj{3}{2}{2}{1} & $1.74\pm0.03$ & $+7.23\pm0.01$ & $0.71\pm0.01$ &
	     $1.31\pm0.01$\\ 
     SO \jj{6}{5}{5}{4} & $0.18\pm0.06$ & $+7.51\pm0.07$ & $0.50\pm0.17$ &
	     $0.10\pm0.03$\\ 
     \noalign{\smallskip}\hline
   \end{tabular}
 \end{table*}

 \begin{table*}
 \caption{Line parameters for the transitions detected in Inter-clump South
     (ICLS)}
 \label{tline_icls}
   \centering
   \begin{tabular}{lcccc}
     \hline\hline\noalign{\smallskip}
	     &$T_\mathrm{MB}$&$V_\mathrm{LSR}$& $\Delta V$    & Line area \\
     Transition   & (K)      & (km~s$^{-1}$)  & (km~s$^{-1}$)&(K km~s$^{-1}$)\\
     \noalign{\smallskip}\hline\noalign{\smallskip}
     CN 1--0 & $0.26\pm0.03$ & $+7.20\pm0.10$ & $0.65\pm0.10$ & $0.10\pm0.01$\\
     \noalign{\smallskip}
     \cthd\ \jj{2}{0,1}{1}{1,2} & $0.25\pm0.03$ & $+7.27\pm0.02$ &
	     $0.37\pm0.04$ & $0.10\pm0.01$\\ 
     \noalign{\smallskip}
     \hdco\ \jj{2}{1,2}{1}{1,1} & $0.90\pm0.07$ & $+7.21\pm0.01$ &
	     $0.46\pm0.03$ & $0.44\pm0.02$\\ 
     \hdco\ \jj{3}{1,3}{2}{1,2} & $0.37\pm0.05$ & $+7.23\pm0.02$ &
             $0.64\pm0.06$ & $0.25\pm0.02$\\
     \hdco\ \jj{3}{1,2}{2}{1,1} & $0.33\pm0.08$ & $+7.24\pm0.02$ &
             $0.38\pm0.06$ & $0.14\pm0.02$\\
     \noalign{\smallskip}
     \htcop\ 1--0 & $0.38\pm0.02$ & $+7.40\pm0.01$ & $0.44\pm0.02$ &
	     $0.18\pm0.01$\\ 
     \htcop\ 2--1 & $0.24\pm0.07$ & $+7.44\pm0.06$ & $0.36\pm0.15$ &
	     $0.09\pm0.03$\\ 
     \noalign{\smallskip}
     SO \jj{3}{2}{2}{1} & $1.99\pm0.03$ & $+7.32\pm0.01$ & $0.47\pm0.01$ &
	     $1.00\pm0.01$\\ 
     \noalign{\smallskip}
     SO$_2$ \jj{3}{1,3}{2}{0,2}	& $0.14\pm0.04$ & $+7.23\pm0.03$ &
	     $0.23\pm0.07$ & $0.04\pm0.01$ \\ 
     \noalign{\smallskip}\hline
   \end{tabular}
 \end{table*}

 \begin{table*}
 \caption{Line parameters for the transitions with hyperfine structure}
 \label{tline_hfs}
   \centering
   \begin{tabular}{llcccc}
     \hline\hline\noalign{\smallskip}
              &	           & $A\tau$ & $V_\mathrm{LSR}$ & $\Delta V$    & \\
     Position & Transition & (K) & (km~s$^{-1}$) & (km~s$^{-1}$) & $\tau$ \\ 
     \noalign{\smallskip}\hline\noalign{\smallskip}
     CL1 & CCH 1--0  & $0.90\pm0.07$ & $+7.19\pm0.01$ & $0.45\pm0.02$ &
         $0.33\pm0.29$ \\ 
        & CCH 2--1   & $0.88\pm0.72$ & $+7.13\pm0.02$ & $0.28\pm0.09$ &
         $1.97\pm2.75$ \\
        & CN 1--0    & $7.76\pm0.83$ & $+7.09\pm0.01$ & $0.35\pm0.01$ &
         $5.90\pm0.83$ \\
        & CN 2--1    & $0.83\pm0.18$ & $+7.00\pm0.01$ & $0.40\pm0.03$ &
         $0.13\pm2.31$\\
        & NO \jf{3}{2}{5}{2}--\jf{1}{2}{3}{2} & $0.78\pm0.11$ & $+7.09\pm0.01$
              & $0.48\pm0.04$ & $0.38\pm0.28$ \\ 
     \noalign{\smallskip}\hline\noalign{\smallskip}
     CL6 & CCH 1--0  & $0.16\pm0.01$ & $+7.24\pm0.02$ & $0.58\pm0.05$ &
              $<0.10$  \\  
         & CN 1--0   & $0.58\pm0.07$ & $+7.24\pm0.01$ & $0.57\pm0.04$ &
              $1.37\pm0.38$ \\ 
         & NO \jf{3}{2}{5}{2}--\jf{1}{2}{3}{2} & $0.36\pm0.03$ &
              $+7.21\pm0.02$ & $0.54\pm0.07$ & $<0.10 $   \\ 
     \noalign{\smallskip}\hline\noalign{\smallskip}
     ICLN & CCH 1--0  & $0.12\pm0.01$ & $+7.29\pm0.02$ & $0.76\pm0.07$ &
              $<0.10$ \\ 
          & CN 1--0   & $0.28\pm0.05$ & $+7.22\pm0.02$ & $0.62\pm0.05$ &
              $1.09\pm0.47$ \\
          & NO \jf{3}{2}{5}{2}--\jf{1}{2}{3}{2} & $0.17\pm0.02$ &
              $+7.20\pm0.04$ & $0.64\pm0.10$  & $0.10^a$\\
     \noalign{\smallskip}\hline\noalign{\smallskip}
     ICLS & CCH 1--0  & $0.22\pm0.06$ & $+7.32\pm0.02$ & $0.50\pm0.05$ &
              $0.22\pm0.65$ \\
          & NO \jf{3}{2}{5}{2}--\jf{1}{2}{3}{2} & $0.18\pm0.03$ &
              $+7.26\pm0.07$ & $0.89\pm0.22$ & $0.10^a$\\
     \noalign{\smallskip}\hline
   \end{tabular}

   \begin{minipage}{10.5cm}
     \begin{list}{}{}
       \item[$^a$] Obtained after assuming optically thin lines.
     \end{list}
   \end{minipage}

 \end{table*}

 \begin{table*}
 \caption{Line intensity $3\sigma$ upper limits for the transitions not
   detected} 
 \label{tline_nd}
   \centering
   \begin{tabular}{lrrrr}
     \hline\hline\noalign{\smallskip}
     & \multicolumn{4}{c}{Position} \\ 
     \cline{2-5}\noalign{\smallskip}
				& CL1         & CL6     &  ICLN   & ICLS \\
     & $T_\mathrm{MB}$ & $T_\mathrm{MB}$ & $T_\mathrm{MB}$ & $T_\mathrm{MB}$ \\
     Transition			& (K)         & (K)     & (K)     & (K)  \\
     \noalign{\smallskip}\hline\noalign{\smallskip}
     CCH 2--1		        &             & $<0.15$	& $<0.14$ & $<0.20$ \\
     CCH 3--2			& $<0.72$     & $<0.58$ & $<0.50$ & $<0.72$ \\
     \noalign{\smallskip}
     CN 2--1		        &             & $<0.14$ & $<0.12$ & $<0.12$ \\
     \noalign{\smallskip}
     CS 5--4		        & $<0.19$     & $<0.34$ & $<0.28$ & $<0.29$ \\
     \noalign{\smallskip}
     \cthd\ \jj{4}{1,4}{3}{0,3} &             & $<0.13$ & $<0.17$ & $<0.19$ \\
     \noalign{\smallskip}
     HCN 3--2		        &             & $<0.60$ & $<0.52$ & $<0.49$ \\
     \noalign{\smallskip}
     \htcop\ 2--1		&      	      & $<0.24$ & $<0.20$ &        \\
     \htcop\ 3--2	        &             &         & $<0.36$ & $<0.37$ \\
     \noalign{\smallskip}
     NO \jf{5}{2}{7}{2}--\jf{3}{2}{5}{2} &    & $<0.43$ & $<0.31$ & $<0.59$  \\
     \noalign{\smallskip}
     SO \jj{4}{5}{3}{4}	        &             & $<0.10$ & $<0.10$ & $<0.22$ \\
     SO \jj{6}{5}{5}{4}	        &	      & $<0.31$ &         & $<0.24$ \\
     SO \jj{7}{6}{6}{5}	        & $<0.87$     & $<0.45$ & $<0.42$ & $<0.49$ \\
     \noalign{\smallskip}
     SO$_2$ \jj{3}{1,3}{2}{0,2} &	      & $<0.14$ & $<0.12$ &   \\
     SO$_2$ \jj{7}{1,7}{6}{0,6} & $<0.26$     & $<0.35$ & $<0.25$ & $<0.30$ \\
     \noalign{\smallskip}\hline
   \end{tabular}
 \end{table*}

\section{Determination of the best-fit solutions using the RADEX modelling}
 \label{radexmethod}

\subsection{Method for multiple-line detected molecules}
 \label{generalradex}

  We used for our calculations the line intensities, $T^i_{mb}$, of \hdco,
  \htcop, and SO, which have at least three observed transitions at each of
  the four positions. We ran a grid of RADEX models using the following ranges
  in temperature, volume density, and molecular column density: $T_\mathrm{K}$
  = 10--30~K; $n$(H$_2)=10^3$--$10^7$\cmt; and
  $N(X)=10^{11}$--$10^{15}$\cmd. For each set of temperature, volume density,
  and column density values, RADEX provides the line intensity for each
  $i$-transition, $T^i_{rx}$. The expected, or calculated, mean beam
  temperature is given by
  \begin{equation}
    T^i_{calc} = T^i_{rx} \eta_{bf}^i
  \end{equation}
  where $\eta_{bf}^i$ is the beam filling-factor of the
  $i$-transition. Assuming a 2-D Gaussian distribution for the source emission
  profile, the filling factor can be expressed as
  \begin{equation}
    \eta_{bf}^i = \frac{\theta_s^2}{\theta_s^2+\theta_{mb}^{i~2}}, 
  \end{equation}
  where $\theta_{mb}^i$ is the FWHM of the antenna main beam at the frequency
  of the $i$-transition, and $\theta_s$ is the FWHM source size.

  We obtained the ``best-fit model'' after searching for the minimum of the
  reduced $\chi^2$ function \citep{Nummelin00}, $\chi^2_\nu$, resulting from
  comparing the measured and calculated line intensities and line intensity
  ratios for the three aforementioned molecules:
  \begin{align}
    \chi^2_{\nu} = \frac{1}{n-p} \sum_{k=1}^3 \biggl[ \left
    (\frac{R^{12k}_{obs} - R^{12k}_{calc}}{\sigma^{r12k}_{obs}} \right )^2 +
    ~~~~~~~~ \\
    \left (\frac{R^{13k}_{obs} - R^{13k}_{calc}}{\sigma^{r13k}_{obs}} \right
    )^2 \nonumber +  \left (\frac{T^{1k}_{obs} -
    T^{1k}_{calc}}{\sigma^{1k}_{obs}} \right )^2  \biggr ]
  \end{align} 
  where $T^{1k}_{obs}$ is the observed line intensity of the lowest frequency
  transition (which is usually the best determined) of the $k$ molecule,
  $R^{ijk}_{obs}$ is the observed line intensity ratio between the $i$- and
  $j$ transitions of the $k$ molecule, $\sigma^{1k}_{obs}$ is the $1\sigma$
  uncertainty associated to the observed line intensity of the lowest
  frequency of the $k$ molecule, $\sigma^{rijk}_{obs}$ is the uncertainty
  associated to $R^{ijk}_{obs}$, $n$ is the number of data points and $p$ is
  the number of free parameters. The $calc$ sub-indexes indicate the
  equivalent values for the line intensities derived with RADEX. For the
  undetected transitions, we only included in our calculations the intensities
  from RADEX that were larger or equal, within errors, than the observed ones.

  Since the real size of the emitting area at each position was unknown, we
  used two sets of values of filling-factors in our analysis:
  $\eta^i_{bf}\simeq 1$, obtained by choosing $\theta_s=1000''$, and
  $\eta^i_{bf}<1$. For the second case, we tested values between 30$''$,
  similar to the beam size at 3-mm, and $50''$, the deconvolved size of the
  CL1/SMM5 dust condensation as well as the size of the CL6 clump derived by
  MGE05.

\subsection{Method for the rest of molecules}
 \label{restmolecules}

  Following a method similar to the one described in Sect.~\ref{generalradex},
  but fixing the values of $n(\mathrm{H_2})$, $T_\mathrm{K}$, and $\theta_s$
  to the best-fit values determined for each position, we looked for the best
  set of values of column density, $N$, excitation temperature, $\Tex$, and
  line opacity, $\tau$ for the rest of molecules.

  We did not follow this method for the CCH lines, because the collisional
  rates for this molecule are still unknown \citep[see e.g.,][]{Padovani09},
  nor for the CN and NO transitions, because the RADEX code does not take into
  account the hyperfine line structure, which in this case ends up
  underestimating the opacity of the lines. Instead, we calculated $\Tex$ and
  $\tau$ of the lines from the simultaneous fitting of the hyperfine
  components. In this last case, we assumed a beam-filling factor, $\eta=1$.

\end{document}